\documentstyle[12pt,fleqn,psfig]{article}
\gdef\journal#1,#2,#3,#4.{{ #1~}{\bf #2}, #3 (#4) }

\def\be{\begin{equation}}\def\bea{\begin{eqnarray}}\def\beaa{\begin{eqnarray*}}
  \def\ee{\end{equation}}  \def\eea{\end{eqnarray}}  \def\eeaa{\end{eqnarray*}}

\def\fun#1#2{\lower3.6pt\vbox{\baselineskip0pt\lineskip.9pt
        \ialign{$\mathsurround=0pt#1\hfill##\hfil$\crcr#2\crcr\sim\crcr}}}
\begin{document}
\def\half{{\textstyle{ 1\over 2}}}
\def\frac#1#2{{\textstyle{#1\over #2}}}
\def\gsim{\mathrel{\raise.3ex\hbox{$>$\kern-.75em\lower1ex\hbox{$\sim$}}}}
\def\lsim{\mathrel{\raise.3ex\hbox{$<$\kern-.75em\lower1ex\hbox{$\sim$}}}}
\def\la{\bigl\langle} \def\ra{\bigr\rangle}
\def\cd{\!\cdot\!}
\def\a{\hat a}      \def\b{\hat b}      \def\c{\hat c}
\def\ab{\a\cd\b}    \def\ac{\a\cd\c}    \def\bc{\b\cd\c}
\def\cg{\cos\gamma} \def\ca{\cos\alpha} \def\cb{\cos\beta}
\def\gg{\hat\gamma}    \def\go{ \hat\gamma_1}
\def\gt{\hat\gamma_2}  \def\gth{\hat\gamma_3}
\def\gf{\hat\gamma_4}

\def\got{ \hat\gamma_1\cd\hat\gamma_2} \def\ggo{ \hat\gamma\cd\hat\gamma_1}
\def\goth{\hat\gamma_1\cd\hat\gamma_3} \def\ggt{ \hat\gamma\cd\hat\gamma_2}
\def\gtth{\hat\gamma_2\cd\hat\gamma_3} \def\ggth{\hat\gamma\cd\hat\gamma_3}
\def\gof{ \hat\gamma_1\cd\hat\gamma_4} \def\ggf{ \hat\gamma\cd\hat\gamma_4}
\def\gtf{ \hat\gamma_2\cd\hat\gamma_4}
\def\gthf{\hat\gamma_3\cd\hat\gamma_4}

\def\n{\hat n}       \def\no{\hat n_1}   \def\nt{\hat n_2}  \def\nth{\hat n_3}
\def\nont{\no\cd\nt} \def\nonth{\no\cd\nth} \def\ntnth{\nt\cd\nth}

\def\nogo{\no\cd\hat\gamma_1} \def\nogt{\no\cd\hat\gamma_2}
\def\nogth{\no\cd\hat\gamma_3}
\def\ntgo{\nt\cd\hat\gamma_1} \def\ntgt{\nt\cd\hat\gamma_2}
\def\ntgth{\nt\cd\hat\gamma_3}
\def\nthgo{\nth\cd\hat\gamma_1} \def\nthgt{\nth\cd\hat\gamma_2}
\def\nthgth{\nth\cd\hat\gamma_3}
\def\D{ {\Delta T \over T} }   \def\dO{d\Omega}



\begin{flushright}
{\footnotesize
SISSA REF. 107/94/A\\
CfA-3915}
\end{flushright}

\vspace{0.1in}

\renewcommand{\thefootnote}{\fnsymbol{footnote}}

\begin{center}
{\Large {\bf Cosmic Strings }}

\vspace{.13in}

{\Large {\bf \& }}

\vspace{.13in}

{\Large {\bf Cosmic Variance }}

\vspace{.3in}

{\bf Alejandro Gangui}$^1$
\footnote{Electronic mail: {\tt gangui@tsmi19.sissa.it}}
and
{\bf Leandros Perivolaropoulos}$^2$
\footnote{Electronic mail: {\tt leandros@cfata3.harvard.edu}}
\footnote{Address after September 1, 1994: Department of Physics,
M.I.T., Cambridge, MA 02139}
\vspace{.13in}

$^1${\em SISSA -- International School for Advanced Studies, \\
via Beirut 2 -- 4, 34013 Trieste, Italy.}\\

\vspace{.06in}

$^2${\em Division of Theoretical Astrophysics \\
         Harvard-Smithsonian Center for Astrophysics \\
                 60 Garden St. \\
            Cambridge, Mass. 02138, USA. }\\

\vspace{.3in}

Submitted to {\em The Astrophysical Journal}


\end{center}

\vfill

\begin{center}
\section*{Abstract}
\end{center}

By using a simple analytical model based on counting random multiple
impulses inflicted on photons by a network of cosmic strings we
show how to construct the general q-point temperature
correlation function of the
Cosmic Microwave Background radiation.
Our analysis is sensible specially for large angular scales
where the Kaiser-Stebbins effect is dominant. Then we concentrate our
study on the four-point function and in particular on its zero-lag limit,
namely, the excess kurtosis parameter, for which we obtain a predicted value
of $\sim 10^{-2}$.
In addition, we estimate the cosmic variance for the kurtosis due to a
Gaussian fluctuation field,
showing its dependence on the primordial spectral index of
density  fluctuations $n$ and finding agreement with previous published results
for the  particular case of a flat Harrison-Zel'dovich spectrum.
Our value for the kurtosis compares well with previous analyses
but falls below the threshold imposed by the cosmic variance when
commonly accepted parameters from string simulations are considered.
In particular the non-Gaussian signal is found to be inversely
proportional to the scaling number of defects, as could be expected by
the central limit theorem.

\vspace{0.7in}

{\it Subject headings:} cosmic microwave background - cosmic strings


\renewcommand{\thefootnote}{\arabic{footnote}}
\addtocounter{footnote}{-2}

\newpage


\vspace{18pt}
\section{Introduction}

A central concept for particle physics theories attempting to unify the
fundamental interactions is the concept of symmetry breaking.
This symmetry breaking
plays a crucial role in the Weinberg-Salam standard electroweak
model ( Masiero 1984) whose extraordinary success in explaining electroweak
scale physics reaches a  precision rarely found before in other areas of
science
(Koratzinos 1994).
In the context of
the standard Big Bang cosmological theory the spontaneous breaking of
fundamental symmetries is realized as a phase transition in the early
universe.
Such phase transitions have several exciting cosmological consequences thus
providing an important link between particle physics and cosmology.

A particularly interesting cosmological issue
is the origin of
structure in the universe. This structure is believed to have emerged by the
gravitational growth of primordial matter fluctuations which are superposed on
the smooth background required by the cosmological principle, the main
assumption of the Big Bang theory.

The above mentioned link of cosmology to particle physics theories has led to
the generation of two classes of theories which attempt to provide physically
motivated solutions to the problem of the origin of structure in the universe.
According to the one class of theories, based on inflation,
primordial fluctuations arose from zero point quantum fluctuations
of a scalar field during an
epoch of superluminal expansion of the scale factor of the universe.
These fluctuations may be shown to obey Gaussian statistics to a very
high degree and to have an approximately scale invariant power spectrum.

According to the second class of theories, those based on topological defects,
primordial fluctuations were produced by a superposition of seeds made of
localized energy density trapped during a symmetry breaking phase transition in
the early universe. Topological defects with linear geometry are known as
cosmic strings and may be shown to be consistent with standard cosmology
unlike
their pointlike (monopoles) and planar (domain walls)
counterparts which require dilution by inflation to avoid overclosing the
universe. Cosmic strings are predicted to form during a phase transition in
the early universe by many but not all Grand Unified Theories (GUTs).

The main elegant feature of the cosmic string model that has caused
significant attention during the past decade is that the only free parameter of
the model (the effective mass per unit length of the wiggly string $\mu$ )
is fixed to approximately the same value from two completely independent
directions.
{}From the {\it microphysical} point of view the constraint
$G\mu\simeq (m_{GUT}/m_P)^2 \simeq 10^{-6}$  is
imposed in order that strings form during the physically realizable GUT phase
transition.
{}From the {\it macrophysical} point of view the same constraint
arises by demanding that the string model be consistent
with measurements of Cosmic Microwave Background (CMB) anisotropies and that
fluctuations are strong enough for structures to form by the present time.
This mass-scale ratio for $G\mu$ is actually a very attractive feature
and even models of inflation have been proposed
(Freese et al. 1990, Dvali et al. 1994)
where observational predictions are related to similar mass scale
relations.

Cosmic strings can account for the formation of large scale
filaments and sheets
(Vachaspati 1986;
Stebbins et al. 1987; Perivolaropoulos, Brandenberger \& Stebbins 1990;
Vachaspati \& Vilenkin 1991; Vollick 1992; Hara \& Miyoshi 1993),
galaxy formation at epochs $z\sim 2-3$
(Brandenberger et al. 1987) and galactic
magnetic fields (Vachaspati 1992b).
They also generate peculiar velocities on large scales
(Vachaspati 1992a; Perivolaropoulos \& Vachaspati 1993),
and are consistent with the amplitude, spectral index
(Bouchet, Bennett \& Stebbins 1988; Bennett, Stebbins \& Bouchet 1992;
Perivolaropoulos 1993a; Hindmarsh 1993)
and the statistics (Gott et al. 1990; Perivolaropoulos
1993b; Moessner, Perivolaropoulos \& Brandenberger 1993;
Coulson et al. 1993; Luo 1994)
of the CMB anisotropies measured by COBE
on angular scales of order $\theta\sim 10^\circ$.

Strings may also leave their imprint on the CMB mainly in three different ways.
The best studied mechanism for producing temperature fluctuations on the
CMB by cosmic strings is the Kaiser-Stebbins effect
(Kaiser \&  Stebbins 1984; Gott 1985).
According to this effect, moving long strings present between the time
of recombination $t_{rec}$ and today produce
(due to their deficit angle (Vilenkin 1981))
discontinuities in the CMB temperature between photons reaching the
observer through opposite sides of the string.
Another mechanism for producing CMB fluctuations by cosmic strings is based
on potential fluctuations on the last scattering surface (LSS).
Long strings and loops present between the time of equal matter and
radiation $t_{eq}$ and the time of recombination $t_{rec}$ induce
density and  velocity fluctuations to their surrounding matter.
These fluctuations grow gravitationally and at $t_{rec}$ they
produce potential fluctuations on the LSS.
Temperature fluctuations arise because photons have to climb out of a potential
with spatially dependent depth.
A third mechanism for the production of temperature anisotropies is based on
the Doppler effect. Moving long strings present on the LSS drag
the surrounding plasma and produce velocity fields.
Thus, photons scattered for
last time on these perturbed last scatterers suffer temperature
fluctuations due to the Doppler effect.

It was recently shown (Perivolaropoulos 1994)
how, by superposing the effects
of these three mechanisms at all times from $t_{rec}$ to today, the power
spectrum of the total temperature perturbation
may be obtained. It turns out that (assuming standard recombination) both
Doppler and potential fluctuations at the LSS dominate over
post-recombination effects on angular scales below $2^\circ$.
However this is not the case for very large scales (where we will be
focusing in the present paper) and this justifies our neglecting the
former two sources of CMB anisotropies. The main effect of these neglected
perturbations is an increase of the gaussian character of the fluctuations
on small angular scales.

The main assumptions of the model were explained in (Perivolaropoulos 1993a).
Here we will only review them briefly for completeness.
As mentioned above, discontinuities in the temperature of the photons
arise due to the peculiar nature of the spacetime
around a long string which even though is {\it locally} flat, {\it globally}
has the geometry of a cone with deficit angle $8\pi G\mu$.
Several are the cosmological effects produced by the mere existence of
this deficit angle (Shellard 1994),
e.g., arcsecond-double images from GUT strings at
redshifts $z\sim 1$, flatten structures from string wakes or elongated
filamentary structures from slowly moving long wiggly strings and, of more
relevance in our present work, post-recombination CMB anisotropy (White, Scott
and Silk 1994) string induced effects (Kaiser \& Stebbins 1984).

The magnitude of the discontinuity is proportional not only to the deficit
angle but also to the string velocity $v_s$ and depends on the relative
orientation between the unit vector along the string ${\hat s}$ and
the unit photon wave-vector ${\hat k}$.
It is given by (Stebbins 1988)
\begin{equation}
{{\Delta T}\over T}=\pm 4\pi G\mu v_s \gamma_s {\hat k}
   \cdot ({\hat v_s}\times {\hat s})
\end{equation}
where $\gamma_s$ is the relativistic Lorentz factor and the sign changes when
the string is crossed.
Also, long strings within each horizon have random velocities, positions
and orientations.

We discretize the time between $t_{rec}$ and today by a set of $N$ Hubble
time-steps $t_i$ such that $t_{i+1} = t_i \, \delta $, i.e., the horizon gets
multiplied by $\delta $ in each time-step.
For $z_{rec}\sim 1400$ we have $N \simeq \log_{\delta}[(1400)^{3/2}]$.

In the frame of the multiple impulse approximation the effect of the
string network on a photon beam is just the linear superposition of
the individual effects, taking into account compensation
(Traschen et al. 1986; Veeraraghavan \& Stebbins 1990; Magueijo 1992),
that is, only those strings within a horizon distance from the beam
inflict perturbations to the photons.

In the following section we show how to construct the general q-point
function of CMB anisotropies at large angular scales produced through the
Kaiser-Stebbins effect. Explicit calculations are performed for the
four-point function and its zero-lag limit, the kurtosis.
Next, we calculate the (cosmic) variance for the kurtosis assuming Gaussian
statistics for arbitrary value of the spectral index and compare it with
the string predicted value (section 3).
Finally, in section 4 we briefly discuss our results.

\vspace{18pt}
\section{The Four--Point Temperature Correlation Function}

According to the previous description, the total temperature shift
in the $\gg$ direction due to the
Kaiser-Stebbins effect on the microwave photons between the time of
recombination and today may be written as
\be
\label{1}
\D(\gg) = 4\pi G \mu v_s \gamma_s
\sum_{n=1}^{N}\sum_{m=1}^{M} \beta^{mn}(\gg)
\ee
where $\beta^{mn}(\gg)$ gives us information about the velocity
$v^{mn}$ and orientation $s^{mn}$
of the $m$th string at the $n$th Hubble time-step and
may be cast as
$\beta^{mn}(\gg) =  \gg\cdot\hat R^{mn} $, with
$\hat R^{mn} = v^{mn} \times s^{mn} =
( \sin\theta^{mn}\cos\phi^{mn} ,
  \sin\theta^{mn}\sin\phi^{mn} ,
  \cos\theta^{mn}                ) $
a unit vector whose direction varies randomly according to the also
random  orientations and velocities in the string network.
In Eq.(\ref{1}), $M$ denotes the mean number of strings per horizon
scale, obtained from simulations ( Allen \& Shellard 1990, Bennett \&
Bouchet 1988) to be of order $M \sim 10$.

Let us now study the correlations in temperature anisotropies.
We are focusing here in the stringy-perturbation inflicted on photons
after the time of recombination on their way to us.
Photons coming from different directions of the sky share a common
history that may be long or short depending on their angular separation.
We therefore take
$\theta_{t_p} \simeq z(t_p)^{-1/2}$
to be the angular size of the horizon at
Hubble time-step $t_p$ ($1\leq p \leq N$).
The kicks on two different photon beams separated by an angular scale
$\alpha_{12} = \arccos(\got)$
greater than $\theta_{t_p}$ will be uncorrelated  for the time-step $t_p$
(no common history up until $\sim t_p$)
but will eventually become correlated (and begin sharing a common past)
afterwards when the horizon increases, say,
at $t_{p'}$ when $\theta_{t_{p'}} \gsim \alpha_{12}$, i.e., when $\alpha_{12}$
{\it fits} in the horizon scale.

Much in the same way, kicks inflicted on three photon beams will be
uncorrelated if at time $t$ any one of the three angles between any two
directions (say, $\alpha_{12}$, $\alpha_{23}$, $\alpha_{13}$)
is greater than $\theta_{t}$, the size of the horizon at time
$t$.
So, in this case, we will be summing (cf. Eq.(\ref{1}))
over those Hubble time-steps $n$
greater than $p$, where $p$ is the time-step when the condition
$\theta_{t_p} = {\rm Max}[\alpha_{12}, \alpha_{23}, \alpha_{13}]$ is satisfied.
The same argument could be extended to any number of photon beams (and
therefore for the computation of the q--point function of temperature
anisotropies)\footnote{A general expression (suitable to whatever angular
scale and to whatever source of temperature fluctuations) for the
three-point correlation function was given in (Gangui, Lucchin, Matarrese
and Mollerach 1994). Although of
much more complexity, similar analysis may be done for the four-point
function and a completely general expression in terms of the angular
trispectrum may be found (Gangui \& Perivolaropoulos 1994).}.

Let us put the above considerations on more quantitative grounds, by
writing the mean q-point correlation function.
Using Eq.(\ref{1}) we may express it as
\be
\label{qpoint}
<{{\Delta T}\over T}(\hat\gamma_1) ... {{\Delta T}\over T}(\hat\gamma_q)>
=
\xi^q \sum_{n_1,...,n_q=1}^N
\sum_{m_1,...,m_q=1}^M <\beta_1^{m_1 n_1} ... \beta_q^{m_q n_q}>
\ee
where $\hat\gamma_1 ...\hat\gamma_q$
are unit vectors denoting directions in the sky
and where $\beta_j^{m_j n_j}= {\hat\gamma_j}\cdot {\hat R^{m_j n_j}}$ with
${\hat\gamma_j}=
(\sin \theta_j \cos \phi_j,\sin\theta_j \sin \phi_j, cos\theta_j)$.
In the previous equation we defined
$\xi \equiv 4 \pi G \mu v_s \gamma_s $.
For simplicity we may always choose a coordinate system such that
$\theta_1=0$, $\phi_1 = 0$ (i.e., $\hat\gamma_1$ lies on the $\hat z$ axis)
and $\phi_2 = \pi/2$.
The seemingly complicated sum (\ref{qpoint}) is in practice
fairly simple to calculate because of the large number of terms that vanish
due to lack of correlation, after the average is taken. The calculation
proceeds by first splitting the product
$\beta_1^{m_1 n_1}...\beta_q^{m_q n_q}$ into all
possible sub-products that correspond to correlated kicks
(i.e., $\beta^{m_j n_j}$'s with the same pair ($m_j, n_j$))
at each expansion Hubble-step and then evaluating the ensemble
average of each sub-product by integration over all directions of ${\hat R}$.
To illustrate this technique we evaluate the two, three and four-point
functions below.

The calculation of {\it the two-point function} has been performed in
(Perivolaropoulos 1993a) but we briefly repeat it here for clarity and
completeness.
Having a correlated pair of beams in  $\go$ and $\gt$ directions from a
particular time-step $p$ onwards means simply that
$\hat R^{m_1 n_1} = \hat R^{m_2 n_2} \Leftrightarrow
n \equiv n_1 = n_2 > p ,\, m \equiv m_1 = m_2$; otherwise the
$\hat R$'s remain uncorrelated and there is no contribution to the mean
two-point function.
Therefore we will have
\be
\label{3}
\D(\gg_1) \D(\gg_2) = \xi^2
\sum_{n=p}^{N}\sum_{m=1}^{M} (\go\cdot\hat R^{mn})(\gt\cdot\hat R^{mn})
\ee
where we wrote just the
correlated part on an angular scale $\alpha_{12} = \arccos(\got)$.
The uncorrelated part on this scale will vanish when performing the
ensemble average $\la \cdot \ra$.
Thus the mean two-point function may be written as
\be
<{{\Delta T}\over T}({\hat \gamma_1}){{\Delta T}\over T}({\hat \gamma_2})>=
\xi^2 <\beta_1 \beta_2> N_{cor} (\alpha_{12})
\ee
where
\be
N_{cor}(\alpha_{12})
\equiv M
[N - 3 \log_{\delta} ( 1 + { \alpha_{12}  \over  \theta_{t_{rec}} } ) ]
\ee
%
is the number of correlated `kicks' inflicted on a scale $\alpha_{12}$
after this scale enters the horizon.
$\theta_{rec}$ is the angular scale of the horizon at
the time of recombination.
Since we may always take
\be
\beta_1 = (0,0,1)\cdot {\hat R}\\
\beta_2 = (\sin \alpha_{12}, 0, \cos \alpha_{12}) \cdot {\hat R}
\ee
with
${\hat R}=(\sin \theta \cos \phi,\sin\theta \sin \phi, cos\theta)$,
$<{{\Delta T}\over T}({\hat\gamma_1}){{\Delta T}\over T}({\hat\gamma_2})>$
may be calculated  by integrating over
$(\theta, \phi)$ and dividing by $4\pi$. The result is
\be
<{{\Delta T}\over T}({\hat\gamma_1}){{\Delta T}\over T}({\hat\gamma_2})>
= \xi^2 {{\cos \alpha_{12}}\over 3} N_{cor}(\alpha_{12})
\ee
which may be shown (Perivolaropoulos 1993a) to lead to a slightly tilted scale
invariant spectrum on large angular scales.

The {\it three-point function} may be obtained in a similar way.
However, the superimposed kernels of the distribution turn out to be
symmetric with respect to positive and negative perturbations and
therefore no mean value for the three-point correlation function arises
(in particular also the skewness is zero).
On the other hand the four-point function is easily found and a mean value
for the excess kurtosis parameter (Gangui 1994b) may be predicted,
as we show below.

In the case of the {\it four-point function} we could find terms
where (for a particular time-step) the photon beams in
directions $\go$, $\gt$ and $\gth$ are all correlated amongst themselves
but not the beam $\gf$, which may be taken sufficiently far apart from the
other three directions.
In such a case we have
\be
\label{2}
\la \D(\gg_1) \D(\gg_2) \D(\gg_3) \D(\gg_4) \ra \longrightarrow
\la \D(\gg_1) \D(\gg_2) \D(\gg_3) \ra \,\,  \la \D(\gg_4) \ra
\ee
and (cf. (Perivolaropoulos 1993a)) this contribution vanishes.

Another possible configuration we could encounter is the one in which the
beams are correlated two by two but no correlation exists between
the pairs (for one particular time-step).
This yields three distinct possible outcomes, e.g.,
$\la \D(\gg_1) \D(\gg_2) \ra \, \la \D(\gg_3) \D(\gg_4) \ra $,
and the other two obvious combinations.

The last possibility is having all four photon beams fully correlated
amongst themselves and this yields
$\la \D(\gg_1) \D(\gg_2) \D(\gg_3) \D(\gg_4) \ra_c $,
where the subscript $c$ stands for the {\it connected} part.

In a way completely analogous to that for the two-point function,
the correlated part for the combination of four beams gives
\be
\label{4}
\D(\gg_1) \D(\gg_2) \D(\gg_3) \D(\gg_4) = \xi^4
\sum_{n=p}^{N}\sum_{m=1}^{M}
(\go \cd\hat R^{mn})(\gt\cd\hat R^{mn})
(\gth\cd\hat R^{mn})(\gf\cd\hat R^{mn})
\ee

Now we are ready to write the full mean four--point function as
\bea
\label{5}
\lefteqn{
\la \D(\gg_1) \D(\gg_2) \D(\gg_3) \D(\gg_4) \ra
         }
\nonumber \\
&  &
=\!\! \left[\!
\la \D(\gg_1) \D(\gg_2) \ra \la \D(\gg_3) \D(\gg_4)\ra\! +\!2 {\rm terms}
\right]\!\! + \!\! \la \D(\gg_1) \D(\gg_2) \D(\gg_3) \D(\gg_4) \ra_c
\nonumber \\
&  &
=  3 \left[ {1\over 3} \xi^2 N_{cor}(\theta) \cos(\theta) \right]^2  +
\la \D(\gg_1) \D(\gg_2) \D(\gg_3) \D(\gg_4) \ra_c
\eea
where for simplicity we wrote this equation using the same scale $\theta$
for all directions on the sky (first term after the equality sign);
after all, we will be
interested in the zero lag limit, in which case we will have
$\theta \to 0$.
By using {\it Mathematica} (Wolfram 1991) it is simple to find that
the second term includes just the sum of twenty-one combinations of
trigonometric functions depending on
$\theta_i,\phi_i$, with $i=1,2,3,4$, the spherical angles for
the directions $\gg_i$ on the sky (cf. Eq.(\ref{4})).
These terms are the only ones which
contribute non-vanishingly after the integration over the angles
$\theta^{mn},\phi^{mn}$ is performed (assuming ergodicity).

When we take the zero lag limit (i.e., aiming for the kurtosis)
the above expression gets largely simplified. After normalizing
by the squared of the variance
$\sigma^4 = [{1\over 3} \xi^2 N_{cor}(0)]^2$
and subtracting the disconnected part we
get the excess kurtosis parameter
\be
\label{6}
{\cal K} = {1\over \sigma^4}
\la \D^4(\gg_1) \ra   - 3
= {9\over 5 M N} \simeq (1.125^{~ + 0.675}_{~ - 0.458})\times 10^{-2}
\ee
where we took $M\simeq 10$ and $\delta = 2$ implying $N\simeq 16$.
Quoted errors take into account possible variation of $\delta$ in the range
$1.5 \leq \delta \leq 3$.

Note  that for values of the scaling solution $M$ increasing (large
number of seeds) the actually small non-Gaussian signal
${\cal K}$ gets further depressed, as it could be expected from the
Central Limit Theorem.

\vspace{18pt}
\section{The r.m.s. excess kurtosis of a Gaussian field}

The previous section was devoted to the computation of ${\cal K}$, the
excess kurtosis parameter, as predicted by a simple analytical model in the
framework of the Cosmic String scenario.
We might ask whether this particular non-Gaussian signal has any chance
of actually being unveiled by current anisotropy experiments.
Needless to say, this could provide a significant probe of the structure
of the relic radiation and furthermore give us a hint on the possible
sources of the primordial perturbations that, after non-linear evolution,
realize the large scale structure of the universe as we currently see it.

However, as it was realized some time ago
(Scaramella \& Vittorio 1991; Abbott \& Wise 1984; Srednicki 1993;
Gangui 1994a),
the mere detection of a
non-zero higher order correlation function (e.g., the four-point function)
or its zero lag limit (e.g.,  the kurtosis)
cannot be directly interpreted as a signal for intrinsically non-Gaussian
perturbations.
In order to tell whether or not a particular measured value for the
kurtosis constitutes a significant evidence of a departure from Gaussian
statistics, we need to know the amplitude of the non-Gaussian pattern
produced by a {\it Gaussian} perturbation field. Namely, we need to know
the r.m.s. excess kurtosis of a Gaussian field.

Let us begin with some basics. Let us denote the kurtosis
$ K \equiv \int {\dO_{\gg}\over 4\pi} \D^4(\gg)$
and assume that the underlying statistics is Gaussian,
namely, that the
multipole coefficients $a_\ell^m$ are Gaussian distributed.
Thus, the ensemble average for the kurtosis is given by the well-known
formula: $\la K \ra = 3 \sigma^4$, where $\sigma^2$ is the mean two--point
function at zero lag, i.e., the CMB variance as given by
\be
\label{10}
\sigma^2 \equiv \la C_2 (0) \ra =
{1 \over 4\pi} \sum_{\ell} {(2\ell + 1) \over 5 }{\cal Q}^2{\cal C}_{\ell}
{\cal W}^2_\ell
\ee
where the ${\cal C}_\ell$ coefficients
(here normalized to ${\cal C}_{\ell = 2} = 1$)
are also dependent on the value
for the primordial spectral index of density fluctuations $n$ and are
given by the usual expression in terms of Gamma functions
(Bond \& Efstathiou 1987; Fabbri, Lucchin \& Matarrese 1987)
\be
\label{cl}
{\cal C}_\ell
=
{\Gamma (\ell+\frac{n}{2}-\frac{1}{2})
\Gamma\left(\frac{9}{2}-\frac{n}{2}\right)
\over \Gamma\left(\ell+\frac{5}{2}-\frac{n}{2}\right)
\Gamma (\frac{3}{2}+\frac{n}{2})}
\ee
${\cal Q} = \la Q_2^2 \ra^{1/2}$ is the {\em rms} quadrupole,
simply related to the quantity $Q_{rms-PS}$ defined in
(Smoot et al. 1992; Bennett et al. 1994)
by ${\cal Q} =\sqrt{4\pi}
Q_{rms-PS}/T_0$, with mean temperature $T_0 = 2.726\pm 0.01 K$
(Mather et al. 1994).
In the previous expression ${\cal W}_\ell$ represents the window function
of the specific experiment. Setting ${\cal W}_0 = {\cal W}_1=0$ automatically
accounts for both monopole and dipole subtraction.
In the particular case of the COBE experimental setup we have,
for $\ell \geq 2$,
${\cal W}_{\ell}\simeq\exp\left[-\frac12 \ell({\ell}+1)(3.2^\circ)^2\right]$,
where $3.2^\circ$ is the dispersion of the antenna--beam profile, which
measures the angular response of the detector (e.g. Wright et al. 1992).
Sometimes the quadrupole term is also subtracted from the maps and
in that case we also set ${\cal W}_2=0$.

However, $\la K \ra$ is just the mean value of the distribution and
therefore we cannot know its real value but within some error bars.
In order to find out how probable it is to get this value after a set of
experiments (observations) is performed, we need to know the variance of
the distribution for $K$.
In other words, we ought to know how peaked the
distribution is around its mean value $\la K \ra$.
The width of the distribution is commonly parameterized by
what is called the cosmic variance of the kurtosis
\be
\sigma_{CV}^2 = \la K^2 \ra - \la K \ra^2
\ee
It is precisely this quantity what attaches
theoretical
error bars to the actual value
for the kurtosis.
Therefore, we may heuristically express
the effect of $\sigma_{CV}^2$ on  $K_{Gauss}$  as follows:
$K_{Gauss} \simeq \la K \ra \pm \sigma_{CV}$,  at one sigma level
(a good approximation in the case of a narrow peak).
Re-arranging factors we may write this expression in a way
convenient for comparing it with ${\cal K}$ as follows
\be
\label{11}
{\cal K}_{CV} = {K_{Gauss}\over \sigma^4}-3 \simeq
\pm {\sigma_{CV}\over \sigma^4 }
\ee
where ${\cal K}_{CV}$ is the excess kurtosis parameter (assuming Gaussian
statistics) purely due to the cosmic variance.
Not only is ${\cal K}_{CV}$ in general non-zero, but its magnitude
increases with the theoretical uncertainty ( $\sigma_{CV}$ ) due
to the limitation induced by our impossibility of making measurements in
more than one Universe.

This gives a fundamental threshold that must be overcome by any
measurable kurtosis parameter in order for us to be able to distinguish the
primordial non-Gaussian signal from the theoretical noise in which it is
embedded.
In other words, unless our predicted value for ${\cal K}$
{\it exceeds} ${\cal K}_{CV}$ we will not be able to tell confidently that any
measured value of  ${\cal K}$ is due to primordial non-Gaussianities.

Now that we know the expression for the cosmic variance of the kurtosis,
let us calculate
it explicitly. We begin by calculating $\la K^2 \ra$ as follows
\be
\label{12}
\la K^2 \ra =
\int {\dO_{\gg_1}\over 4\pi} \int {\dO_{\gg_2}\over 4\pi}
\la \D^4(\gg_1) \D^4(\gg_2) \ra
\ee
By assuming Gaussian statistics for the temperature perturbations
$\D(\gg)$ we may make use of standard combinatoric relations, and get
\bea
\label{13}
\lefteqn{
\la \D^4(\gg_1) \D^4(\gg_2) \ra  = \,
9 \, \la\D^2(\gg_1)\ra^2 \la\D^2(\gg_2)\ra^2
         }
\nonumber \\
&  &
+ \, 72 \, \la\D^2(\gg_1)\ra \la\D^2(\gg_2)\ra \la\D(\gg_1)\D(\gg_2)\ra^2
+ \, 24 \, \la\D(\gg_1)\D(\gg_2)\ra^4
\eea
As we know, the ensemble averages are rotationally invariant and
therefore
$\la\D^2(\gg_1)\ra =  \la C_2 (0) \ra \equiv \sigma^2 $
is independent of the direction $\gg_1$.
Plug this last equation into Eq.(\ref{12}) and we get
\be
\label{14}
\la K^2 \ra = 9\sigma^8
     +  36\sigma^4  \int^1_{-1}d\cos\alpha \, \la C_2(\alpha) \ra^2
     +  12 \int^1_{-1}d\cos\alpha \, \la C_2(\alpha) \ra^4 .
\ee
The above integrals may be solved numerically. Then, using this result
in the expression for $\sigma_{CV}^2$ we get ${\cal K}_{CV}$, the value
for the excess kurtosis parameter of a Gaussian field.

It is also instructive to look at Eq.(\ref{14}) in some more detail, so
that the actual dependence on the spectral index becomes clear.
By expanding the mean two-point correlation functions
within this expression in terms of spherical harmonics and after some
long but otherwise straightforward algebra we find
\bea
\label{37}
\lefteqn{
{\cal K}_{CV} =
\left[ \,
72 \, {
\sum_{\ell} (2\ell + 1){\cal C}_{\ell}^2 {\cal W}_\ell^4
\over
[ \sum_{\ell} (2\ell + 1){\cal C}_{\ell} {\cal W}_\ell^2 ]^2}
\right.
}
\nonumber \\
&  &
+ \,
\left.
24 \,
{
\{
\prod_{i=1}^{4}\sum_{\ell_i}\sum_{m_i=-\ell_i}^{\ell_i}
{\cal C}_{\ell_i} {\cal W}_{\ell_i}^2
\}
\left(
\sum_L 4\pi\,
\bar {\cal H}_{\ell_1,\,\,\ell_2\,,\,\,\,\, L}^{m_1, m_2, m_3+m_4}
\bar {\cal H}_{\ell_3,\,\,\ell_4\,,\,\,\,\,\,\, L}^{m_3, m_4, -m_3-m_4}
\right)^2
\over
[ \sum_{\ell} (2\ell + 1){\cal C}_{\ell} {\cal W}_\ell^2  ]^4
}
\right]^{1/2}
\eea
where the coefficients
$\bar {\cal H}_{\ell_1,\,\ell_2,\,\ell_3}^{m_1, m_2, m_3}
\equiv \int \dO_{\gg} {Y_{\ell_1}^{m_1}}(\gg) Y_{\ell_2}^{m_1}(\gg)
Y_{\ell_3}^{m_3}(\gg)$,
which can be easily expressed in terms of
Clebsch--Gordan coefficients (Messiah 1976), are only non--zero if
the indices $\ell_i$, $m_i$ ($i=1,2,3$) fulfill the relations:
$\vert \ell_j - \ell_k \vert \leq  \ell_i \leq \vert \ell_j + \ell_k \vert$,
$\ell_1 + \ell_2 + \ell_3 = even$ and $m_1 + m_2 + m_3 = 0$.
In the above equation the $n$-dependence is hidden inside the multipole
coefficients ${\cal C}_{\ell}$ (cf. Eq.(\ref{cl})).

Eq.(\ref{37}) shows an analytic expression for computing ${\cal K}_{CV}$
which, in turn, represents a fundamental threshold for any given
non-Gaussian signal.
For interesting values of the spectral index (say, between
$0.8\lsim n \lsim 1.3$) we find no important variation in
${\cal K}_{CV}$, being its value consistent with the Monte-Carlo
simulations performed by Scaramella \& Vittorio 1991
(SV91 hereafter) , see below.
These authors concentrated on a Harrison-Zel'dovich spectrum, considered
also the quadrupole contribution and took a slightly different dispersion
width for the window function ($3.0^\circ$ in their simulations).

We coded an IBM RISC 6000 for solving Eq.(\ref{37}) numerically.
We first calculated ${\cal K}_{CV}$ for a somewhat reduced range for the
multipole index $\ell$ with quadrupole subtracted, $3\leq\ell\leq 5$, so
that we were able to test the result with that obtained by using
{\it Mathematica}.
We found perfect agreement and a value
${\cal K}_{CV} \simeq 1.85$ (for $n=1$).
We may even go further and include the quadrupole, i.e.
$2\leq\ell\leq 5$, and in this case we get
${\cal K}_{CV} \simeq 1.92$. This slightly larger value obtained after
including $\ell = 2$ is not a new feature and simply reflects the intrinsic
theoretical uncertainty of the lowest order multipoles; see e.g.
(Gangui et al. 1994) for a similar situation in the case of the skewness.

Of course ${\cal K}_{CV} \simeq 1.85$ (or 1.92 with quadrupole) are still
a factor 6 above the value $\sim 0.3$ found in SV91
(where they included the quadrupole).
Our analytical analysis expresses ${\cal K}_{CV}$ as a ratio of averages
(rather than the average of a ratio as in SV91) and therefore exact
agreement should not be expected.
But still the main reason for the discrepancy lies in the small range for
the multipole index $\ell$.
We therefore increased the value of $\ell_{max}$ and checked that
${\cal K}_{CV}$ monotonically became smaller and smaller.
When we were in the range $3\leq\ell\leq 10$ we got
${\cal K}_{CV} \simeq 1.16$; instead, for $3\leq\ell\leq 15$ we got
${\cal K}_{CV} \simeq 0.98$ (for $n=1$) --clearly a sensible decrease.
CPU--time limitations prevent us from carrying out the numerical
computations for larger values of $\ell_{max}$
(usually $\ell_{max}$ is chosen of order 30, i.e., beyond those values
of $\ell$ where the exponential suppression of the ${\cal W}_{\ell}$
makes higher $\ell$ contribution to the sums negligible), but still we
expect ${\cal K}_{CV}$ to keep on steadily decreasing (we checked that the
decrease in ${\cal K}_{CV}$ was tinier as $\ell_{max}$ got larger).
This, together with some previous experience in similar computations
(Gangui et al. 1994) makes us believe that in the case appropriately large
values for $\ell_{max}$ were used , the SV91 value above mentioned would
be attained (taking into account the quadrupole subtraction, of course).

In addition, we checked that a different value of the spectral index
$n$ does not change the essence of the above considerations.
We explored the cosmologically interesting range $0.8\leq n \leq 1.3$ and
got values $1.22 \geq {\cal K}_{CV} \geq 1.08 $ (taking $3\leq\ell\leq 10$)
and values $  1.05 \geq {\cal K}_{CV} \geq 0.88 $
(taking $3\leq\ell\leq 15$).
We plot ${\cal K}_{CV}$ versus the spectral index for $\ell_{max} = 15$
in Fig.1.

\begin{figure}[t]
 \centerline{\psfig{figure=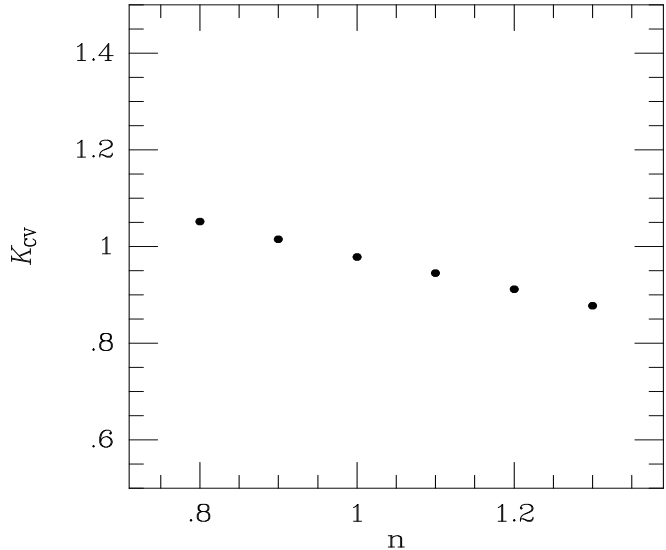,height=4in,width=4in}}
\caption{Excess Kurtosis parameter of a Gaussian temperature fluctuation
 field as function of the spectral index.}
\end{figure}

Note the small rate of variation of ${\cal K}_{CV}$ with $n$ and
that, as expected, ${\cal K}_{CV}$ takes larger values for smaller
spectral indexes.
This is clearly due to the fact that a small $n$ generates more power on
large scales (i.e., small $\ell$) and precisely these scales are the ones
that contribute the most to the cosmic variance of the kurtosis field.

\vspace{18pt}
\section{Discussion}

In the present paper we showed how to implement the multiple impulse
approximation for perturbations on a photon beam
(stemming from the effect of the string network)
in the actual construction of higher order correlations for the CMB
anisotropies.
We then focused on the four-point function and on the excess kurtosis
parameter, finding for the latter a value ${\cal K}\sim 10^{-2}$.

We also calculated explicitly the rms excess kurtosis ${\cal K}_{CV}$
predicted to exist even for a Gaussian underlying field and showed its
dependence on the primordial spectral index of density fluctuations.
This constitutes the main source of theoretical uncertainty at COBE scales.
In fact, the cosmic string signature that might have been observable is
actually blurred in the cosmic variance mist reigning at very large scales.

Nevertheless, there is still a chance of getting a string-characteristic
angular dependence from the study of the mean four-point correlation
function by exploiting the particular geometries deriving from it;
namely, its collapsed cases (where some of the five independent angles are
taken to be zero) or from particular choices for these angles (as in
the case of taking all angles equal).

A preliminary analysis (Gangui \& Perivolaropoulos 1994)
making use of just one non-vanishing angle in a
collapsed configuration as the one mentioned above shows a
potentially interesting effect that could eventually increase
notably the small non-Gaussian signal, and suggests that this is indeed a
subject worth of further investigation.
Some of these alternatives are presently under study
and we expect to report progress on this subject in a future publication.

\vspace{5mm}
\noindent{\bf Acknowledgements:}
It is a pleasure to thank A. Masiero and Q. Shafi for instructive
discussions, and for having made this collaboration possible
by inviting one of us (L.P.) to lecture in the ICTP summer school.
A.G. wants also to acknowledge S. Matarrese and D.W. Sciama for
encouragement, and R. Innocente and L. Urgias for their kind assistance
with the numerical computations.
This work was supported by the Italian MURST (A.G.)
and by a CfA Postdoctoral fellowship (L.P.).

\vspace{18pt}

\section*{References}

\vspace{.2in}

\begin{description}

\item[\rm Abbott, L.F. \& Wise, M.B.] 1984, ApJ, 282, L47.

\item[\rm Allen, B. \& Shellard, E. P. S.] 1990, Phys. Rev. Lett.
{\bf 64}, 119.

\item[\rm Bennett, D. P., Stebbins, A. \& Bouchet, F. R.] 1992, Ap.J.Lett.
{\bf 399}, L5.

\item[\rm Bennett, C.L. el al.] 1994, ApJ, submitted
(preprint astro--ph/9401012).

\item[\rm Bennett, D. P. \& Bouchet, F. R.] 1988, Phys. Rev. Lett.
{\bf 60}, 257.

\item[\rm Bouchet, F. R., Bennett, D. P. \& Stebbins, A.] 1988,
Nature {\bf335}, 410.

\item[\rm Bond, J.R. \& Efstathiou, G.] 1987, MNRAS, 226, 655.

\item[\rm Brandenberger, R., Kaiser, N., Shellard, E. P. S., and Turok, N.]
1987, Phys.Rev. {\bf D36}, 335.

\item[\rm Coulson, D., Ferreira, P., Graham, P. \& Turok, N.] 1993,
{\it $\Pi$ in the Sky? CMB Anisotropies from Cosmic Defects},
PUP-TH-93-1429, HEP-PH/9310322.

\item[\rm  Dvali, G., Shafi, Q., and Schaefer, R] 1994,
preprint: hep-ph/9406319.

\item[\rm Fabbri, R., Lucchin, F. \& Matarrese, S.] 1987, ApJ, 315, 1.

\item[\rm Freese, K., Frieman, J.A., and Olinto, A.V.] 1990,
Phys. Rev. Lett. 65, 3233.

\item[\rm Gangui, A.] 1994a, Phys. Rev. D{\bf 50}, xxx.
(preprint astro-ph/9406014).

\item[\rm Gangui, A.] 1994b, in proceedings of {\it Birth of The Universe
              \& Fundamental Physics}, Rome, May 18 - 21, Ed. F. Occhionero,
              Springer-Verlag, in press.

\item[\rm Gangui, A., Lucchin, F., Matarrese, S. and Mollerach, S.] 1994,
                   Astrophys. J. {\bf 430}, 447.

\item[\rm Gangui, A. and Perivolaropoulos, L.] 1994, work in progress.

\item[\rm Gott, R.] 1985, Ap.J. {\bf 288}, 422.

\item[\rm Gott, J. et al.] 1990, Ap.J. {\bf 352}, 1.

\item[\rm Hara T. \& Miyoshi S.] 1993, Ap. J. {\bf 405}, 419.

\item[\rm Hindmarsh M.] 1993, {\it Small Scale CMB Fluctuations from Cosmic
Strings}, DAMTP-93-17, ASTRO-PH/9307040.

\item[\rm Kaiser, N. \&  Stebbins, A.] 1984, {\it Nature} {\bf 310}, 391.

\item[\rm Koratzinos, M.] invited talk in the ``Workshop on
perspectives in theoretical and experimental particle physics", Trieste,
7-8 July, 1994.

\item[\rm Luo, X.] 1994, ApJ 427, L71.

\item[\rm Magueijo, J.] 1992, Phys. Rev. D{\bf 46}, 1368.

\item[\rm Masiero, A.] 1984,  in {\it Grand Unification with and
without Supersymmetry and Cosmological Implications}, Eds. C. Kounnas et
al, World Scientific; (a comprehensive review of the vices and virtues
of the standard model).

\item[\rm Mather, J. et al.] 1994, ApJ, 420, 439.

\item[\rm Messiah, A.] 1976, {\em Quantum Mechanics}, Vol.2
(Amsterdam: North--Holland).

\item[\rm Perivolaropoulos L.] 1993a, Phys.Lett. {\bf B298}, 305.

\item[\rm Perivolaropoulos L.] 1993b, Phys. Rev. {\bf D48}, 1530.

\item[\rm Perivolaropoulos, L.] 1994,  {\it Spectral Analysis of CMB
Fluctuations Induced by Cosmic Strings}, Submitted to the Ap. J., CfA-3591,
ASTRO-PH/9402024.

\item[\rm Moessner, R., Perivolaropoulos, L. \& Brandenberger, R.] 1994,
Astrophys. J. {\bf 425}, 365.

\item[\rm Perivolaropoulos, L., Brandenberger, R. \& Stebbins, A.] 1990,
Phys.Rev. {\bf D41}, 1764.

\item[\rm Perivolaropoulos, L. \& Vachaspati, T.] 1994, Ap. J. Lett. 423,
L77.

\item[\rm Scaramella, R. \& Vittorio, N.] 1991, ApJ, 375, 439

\item[\rm Shellard, E. P. S.] 1994,
     Lectures presented at SILARG VIII.
     To appear in {\it Gravitation: The spacetime structure}, Eds.
     Letelier P. and Rodrigues W.A., World Scientific.

\item[\rm Smoot, G.F. et al.] 1992, ApJ, 396, L1.

\item[\rm Srednicki, M.] 1993, ApJ, 416, L1.

\item[\rm Stebbins, A.] 1988, Ap. J. {\bf 327}, 584.

\item[\rm Stebbins, A.  et al.] 1987, Ap. J. {\bf 322}, 1.

\item[\rm Traschen, J., Turok, N.,  and Brandenberger, R.] 1986,
Phys. Rev. {\bf D34}, 919.

\item[\rm Vachaspati, T.] 1986, Phys. Rev. Lett. {\bf 57}, 1655.

\item[\rm Vachaspati, T.] 1992a, Phys.Lett. {\bf B282}, 305.

\item[\rm Vachaspati, T.] 1992b, Phys. Rev. D{\bf 45}, 3487.

\item[\rm Vachaspati, T. \& Vilenkin, A.] 1991. Phys. Rev. Lett. {\bf 67},
1057-1061.

\item[\rm Veeraraghavan, S. and Stebbins, A.] 1990,
Astrophys. J. {\bf 365}, 37.

\item[\rm White, M., Scott, D., and Silk, J.] 1994,
Ann. Rev. Astron. and Astrophys., to appear.

\item[\rm Vilenkin, A.] 1981, Phys.Rev. {\bf D23}, 852.

\item[\rm Wolfram, S.] 1991, {\it Mathematica version 2.0},  Addison-Wesley.

\item[\rm Vollick, D. N.] 1992, Phys. Rev. D{\bf 45}, 1884.

\item[\rm Wright, E.L. et al.] 1992, ApJ, 396, L13.

\end{description}



\end{document}